\documentclass[showkeys,aps,showpacs,amssymb,prd]{revtex4}

\usepackage{appendix}
\usepackage{amsbsy}
\usepackage{amsfonts}
\usepackage{amsmath}
\usepackage{amssymb}
\usepackage{amsthm}

\usepackage{graphicx}
\usepackage{ifthen}
\usepackage{textcomp}

\newcounter{multi} \newcounter{multa}

\newcounter{faki} \newcounter{faka}

    \newtheorem{theorem}{Theorem}

\theoremstyle{definition} 

\theoremstyle{remark} 

\def\bbf{{\bf f}}
\def\bc{{\bf c}}
\def\bx{{\bf x}}

\def\bs{{\bf s}}

\def\detfA{{\rm det}\left({\rm Jac}\,\bbf^{A_n}_{\bc}\right)}
\def\detfD{{\rm det}\left({\rm Jac}\,\bbf^{D_{n}^{\pm}}_{\bc}\right)}

\def\detfEseven{{\rm det}\left({\rm Jac}\,\bbf^{E_7}_{\bc}\right)}
\def\detfEeight{{\rm det}\left({\rm Jac}\,\bbf^{E_8}_{\bc}\right)}
\def\fA{{\bbf}^{A_n}_{\bc}}
\def\fD{{\bbf}^{D_{n}^{\pm}}_{\bc}}

\def\fEseven{{\bbf}^{E_7}_{\bc}}
\def\fEeight{{\bbf}^{E_8}_{\bc}}

\def\det {\mathop{\rm det}\nolimits}

\def\Jac {\mathop{\rm Jac}\nolimits}

\def\RR{\mathbb{R}}

\newcommand{\beqa}{\begin{eqnarray}}
\newcommand{\beq}{\begin{equation}}
\newcommand{\eeqa}{\end{eqnarray}}
\newcommand{\eeq}{\end{equation}}

\newcommand{\bo}{\boldsymbol}
\newcommand{\oli}{\overline}

\newcommand{\rideal}{R/(\varphi(x))}
\newcommand{\lra}{\longrightarrow}
\newcommand{\fkM}{\mathfrak{M}}
\newcommand{\fkm}{\mathfrak{m}}

\begin{document}

\title{A Universal Magnification Theorem III. Caustics Beyond Codimension Five}
\author{A. B. Aazami}\email{aazami@math.duke.edu}\affiliation{Department of
Mathematics, Duke University, Science Drive, Durham, NC 27708}
\author{A. O. Petters}\email{petters@math.duke.edu}\affiliation{Departments of
Mathematics and Physics, Duke University, Science Drive, Durham, NC 27708}

\begin{abstract}
In the final paper of this series, we extend our results on magnification invariants to the infinite family of $A_n~(n \geq 2), D_n~(n \geq 4), E_6, E_7, E_8$ caustic singularities.  We prove that for families of general mappings between planes exhibiting any caustic singularity of the $A_n~(n \geq 2), D_n~(n \geq 4), E_6, E_7, E_8$ family, and for a point in the target space lying anywhere in the region giving rise to the maximum number of lensed images (real pre-images), the total signed magnification of the lensed images will always sum to zero.  The proof is algebraic in nature and relies on the Euler trace formula.
\end{abstract}


\maketitle
\section{Introduction}
\label{Introduction}
In papers I and II of this series (Aazami \& Petters 2009 \cite{Aazami-Petters, Aazami-Petters2}), we established a universal magnification theorem for all higher-order caustics up to codimension 5.  It was shown that to each such caustic singularity is associated a magnification sum rule of the form
$$
\sum_{i} \fkM_{i} = 0.
$$
All of these bifurcation sets or big caustics occur in a $n$-parameter space.  Slices of the big caustics give rise to caustic metamorphoses that occur in gravitational lensing (e.g., Blandford 1990 \cite{Blandford90}, Schneider, Ehlers, \& Falco 1992 \cite{Sch-EF}, Petters 1993 \cite{Petters93}, and Petters et al. 2001 \cite[Chaps. 7, 9]{Petters}).  It was discussed in Paper I, using the hyperbolic umbilic in particular, how the above magnification relations may be used for substructure studies of four-image lens galaxies. 

We now extend these results to all higher-order caustic singularities beyond codimension $5$.  These are classified according to Arnold's $A, D, E$ classification of Lagrangian map-germs (Arnold 1973 \cite{Arnold73}).  Thus, we show that for families of general mappings between planes exhibiting {\it any} caustic singularity with corresponding Coxeter-Dynkin diagram of type $A_n~(n \geq 2), D_n~(n \geq 4), E_6, E_7, E_8$, and for a point anywhere in the region of the target space giving rise to the maximum number of lensed images (real pre-images), the total signed magnification is identically zero.
Our proof is algebraic and relies on the Euler trace formula.  Finally, we emphasize that the magnification sum relations are {\it geometric} invariants, being the reciprocals of Gaussian curvatures at critical points.

The outline of this paper is as follows.  In Section~\ref{ADE} we give a brief overview of the $A, D, E$ family of caustic singularities.  In Section~\ref{Theorem} we state our main theorem.  The proof itself is presented in Appendix~\ref{Proof}.

\begin{table}
\footnotesize
\centering
\vskip 6pt
\begin{tabular}{| c | c |}
\hline
& \\
& $F_{\bc,\bs}(x,y) =  \underbrace{\pm x^{n+1} \pm y^2}_{\text{germ}} + \underbrace{c_{n-1}x^{n-1} + \cdots + c_3x^3 + s_2x^2 - s_1x \pm s_2y}_{\text{unfolding terms}}$ \\
~~~~~~{\it \large{A}}$_{n}~~~(n \geq 2)$~~~~~~ & \\
& ${\bf f}_{\bc}(x,y) = \left(\pm(n+1)x^{n} + (n-1)c_{n-1}x^{n-2} + \cdots + 3c_3x^2 \mp 4yx~,~\mp2y\right)$ \\
& \\
\hline
& \\
& $F_{\bc,\bs}(x,y) = \underbrace{x^2y \pm y^{n-1}}_{\text{germ}} + \underbrace{c_{n-2}y^{n-2} + \cdots + c_2y^2 - s_2y - s_1x}_{\text{unfolding terms}}$ \\
~~~~~~{\it \large{D}}$_n~~~(n \geq 4)$~~~~~~ & \\
& ${\bf f}_{\bc}(x,y) = \left(2xy~,~x^2\pm(n-1)y^{n-2} + (n-2)c_{n-2}y^{n-3} + \cdots + 2c_2y\right)$ \\
& \\
\hline
& \\
& $F_{\bc,\bs}(x,y) = \underbrace{x^3 \pm y^4}_{\text{germ}} + \underbrace{c_3xy^2 + c_2y^2 + c_1xy - s_2y - s_1x}_{\text{unfolding terms}}$ \\
~~~~~~{\it \large{E}}$_6$~~~~~~ & \\
& ${\bf f}_{{\bo c}}(x,y) = \left(3x^2 + c_3y^2 +c_1y~,~\pm4y^3 + 2c_3xy + 2c_2y + c_1x\right)$ \\
& \\
\hline
& \\
& $F_{\bc,\bs}(x,y) = \underbrace{x^3 + xy^3}_{\text{germ}} + \underbrace{c_4y^4 + c_3y^3 + c_2y^2 + c_1xy - s_2y - s_1x}_{\text{unfolding terms}}$ \\
~~~~~~{\it \large{E}}$_7$~~~~~~ & \\
& ${\bf f}_{{\bo c}}(x,y) = \left(3x^2 + y^3 + c_1y~,~3xy^2 + 4c_4y^3 + 3c_3y^2 + 2c_2y + c_1x\right)$ \\
& \\
\hline
& \\
& $F_{\bc,\bs}(x,y) = \underbrace{x^3 + y^5}_{\text{germ}} + \underbrace{c_5xy^3 + c_4xy^2 + c_3y^3 + c_2y^2 + c_1xy - s_2y - s_1x}_{\text{unfolding terms}}$ \\
~~~~~~{\it \large{E}}$_8$~~~~~~ & \\
& ~~~~${\bf f}_{{\bo c}}(x,y) = \left(3x^2 + c_5y^3 + c_4y^2 + c_1y~,~5y^4 + 3c_5xy^2 + 2c_4xy + 3c_3y^2 + 2c_2y + c_1x\right)$~~~~ \\
& \\
\hline
\end{tabular}
\caption{For each type of Coxeter-Dynkin diagram listed, indexed by $n$, the second column shows the corresponding universal local forms of the smooth $(n-1)$-parameter family of general functions $F_{{\bo c},\bs}$, along with their $(n-3)$-parameter family of induced general maps $\bbf_{\bo c}$ between planes.  This classification is due to Arnold 1973 \cite{Arnold73}.}
\label{table1}
\end{table}

\section{Higher-Order Caustics of the $A, D, E$ family}
\label{ADE}

We inherit the notation and terminology of Paper II.  To that end, consider a smooth $n$-parameter family $F_{{\bc},{\bs}}(\bx)$ of functions on an open subset of $\mathbb{R}^2$ that induces a smooth $(n-2)$-parameter family of mappings $\bbf_{\bc}(\bx)$ between planes ($n \geq 2)$.  One uses $F_{\bc,\bs}$ to construct a {\it Lagrangian submanifold} that is projected into the space $\{{\bc},{\bs}\} = \mathbb{R}^{n-2} \times \mathbb{R}^2$.  The caustics of $\bbf_{\bc}$ will then be the critical values of the projection  (e.g.,  Golubitsky \& Guillemin 1973 \cite{Gol-G}, Majthay 1985 \cite{Majthay}, Castrigiano \& Hayes 1993 \cite{C-Hayes}, and \cite[pp. 276-86]{Petters}).  These projections are called {\it Lagrangian maps}, and they are differentiably equivalent to  $\bbf_\bc$.

Arnold classified all stable simple Lagrangian map-germs of $n$-dimensional Lagrangian submanifolds by their generating family $F_{\bc,\bs}$ (\cite{Arnold73}, Arnold, Gusein-Zade, \& Varchenko I 1985 \cite[p. 330-31]{AGV1}, and \cite[p. 282]{Petters}).  In the process he found a connection between his classification and the Coxeter-Dynkin diagrams of the simple Lie algebras of types $A_n~(n \geq 2), D_n~(n \geq 4), E_6, E_7, E_8$.  This classification is shown in Table~\ref{table1}.    The singularities in Paper II arose as follows: $A_2$ (fold), $A_3$ (cusp), $D_4^{-}$ (elliptic umbilic), $D_4^{+}$ (hyperbolic umbilic), $A_4$ (swallowtail), $A_5$ (butterfly), $D_5$ (parabolic umbilic), $A_6$ (wigwam), $E_6$ (symbolic umbilic), $D_6^{-}$ ($2^{\rm nd}$ elliptic umbilic), and $D_6^{+}$ ($2^{\rm nd}$ hyperbolic umbilic). 

For the $\bbf_{\bc}$ shown in Table~\ref{table1}, call $\bx \in \RR^2$ a {\it lensed image} (or a {\it real pre-image}\,) of the {\it target point} $\bs \in \RR^2$ if $\bbf_{\bc}(\bx) = \bs$ (in particular, our lensed images are always in $\RR^2$, not $\mathbb{C}^2$).  A point $\bx_i \in \RR^2$ is a lensed image of the target point $\bs \in \RR^2$  if and only if $\bx_i$ is a critical point of $F_{\bc,\bs}$ (relative to a gradient in $\bx$).  Next, we define the {\it magnification} $\fkM(\bx_i;\bs)$ at a critical point $\bx_i$ of the family $F_{{\bo c}, \bs}$ by the reciprocal of the Gaussian curvature at the point $(\bx_i,F_{{\bo c},\bs}(\bx_i))$ in the graph of $F_{{\bo c},\bs}$:
\beq
\label{Gauss}
\fkM({\bx_i; \bs})
= \frac{1}{{\rm Gauss}(\bx_i,F_{{\bc},\bs}(\bx_i))}\cdot
\eeq
This makes it clear that the magnification invariants established in our theorem are {\it geometric} invariants.  In addition, since ${\rm Gauss}(\bx_i,F_{{\bc},\bs}(\bx_i)) = \text{det(Hess}\,F_{\bc,\bs})(\bx_i)$, and since each $\bbf_{\bc}$ in Table~\ref{table1} satisfies $\text{det(Jac}\,\bbf_{\bc}) = \text{det(Hess}\,F_{\bc,\bs})$, we can also express the magnification in terms of $\bbf_{\bc}$:
\beq
\label{Jac}
\fkM (\bx_i; \bs)
= \frac{1}{\det(\Jac \bbf_{\bc})(\bx_i)}\ ,\nonumber
\eeq
where $\bx_i$ is a lensed image of $\bs$ under $\bbf_{\bc}$.  We now proceed to our main theorem.

\section{Statement of Main Theorem}
\label{Theorem}

\begin{theorem}
\label{theorem-main}
For any of the universal, smooth $(n-1)$-parameter family of general functions $F_{{\bo c},\bs}$ {\rm(}or induced general mappings $\bbf_{\bo c}${\rm)} in Table~\ref{table1}, and for any non-caustic point $\bf s$ (light source position) in the indicated region, the following results hold for the magnification $\fkM_i \equiv  \fkM({\bx_i; \bs})$:

\begin{enumerate}
\item $A_n~(n \geq 2)$ obeys the magnification relation in the $n$-image region: $\sum_{i = 1}^{n} \fkM_i = 0,$
\item $D_n~(n \geq 4)$ obeys the magnification relation in the $n$-image region: $\sum_{i = 1}^{n} \fkM_i = 0,$
\item $E_6$ obeys the magnification relation in the six-image region: $\sum_{i = 1}^{6} \fkM_i = 0,$
\item $E_7$ obeys the magnification relation in the seven-image region: $\sum_{i = 1}^{7} \fkM_i = 0,$
\item $E_8$ obeys the magnification relation in the eight-image region: $\sum_{i = 1}^{8} \fkM_i = 0.$
\end{enumerate}
\end{theorem}
\noindent The magnification relations for $A_2$ (fold) and $A_3$ (cusp) are known \cite{Blan-Nar,Sch-EF,Sch-Weiss92,Zakharov,Petters}.  Those for $D_4^{-}$ (elliptic umbilic), $D_4^{+}$ (hyperbolic umbilic), $A_4$ (swallowtail), $A_5$ (butterfly), $D_5$ (parabolic umbilic), $A_6$ (wigwam), $E_6$ (symbolic umbilic), $D_6^{-}$ ($2^{\rm nd}$ elliptic umbilic), and $D_6^{+}$ ($2^{\rm nd}$ hyperbolic umbilic) were discovered recently in \cite{Aazami-Petters, Aazami-Petters2}. 
\vskip 10 pt

\noindent
{\it Remarks.} First, the results of Theorem~\ref{theorem-main} actually apply even when the non-caustic point $\bs$ is not in the maximum number of real pre-images region.  However, pre-images from $\mathbb{C}^2$ will appear, which are unphysical in gravitational lensing.  Second, it is important to point out that Theorem~\ref{theorem-main} is not a direct consequence of the Euler-Jacobi formula, of multi-dimensional residue integral methods, or of Lefschetz fixed point theory because some of the singularities have fixed points at infinity that must be treated invidually. Third, we point out that for $n \geq 6$ there are Lagrangian maps that cannot be approximated by stable Lagrangian map-germs \cite{Arnold73}.

\section{Conclusion}
\label{Conclusion}
The paper presented a theorem about the magnification of lensed images for all caustic singularities appearing in the infinite family of $A_n~(n \geq 2), D_n~(n \geq 4), E_6, E_7, E_8$ caustic singularities.  We proved that for families of general mappings between planes locally exhibiting any caustic singularity of the $A_n~(n \geq 2), D_n~(n \geq 4), E_6, E_7, E_8$ family, and for a target point lying anywhere in the region giving rise to the maximum number of lensed images (real pre-images), the total signed magnification of the lensed images will always sum to zero.  The signed magnifications are geometric invariants as they are Gaussian curvatures at critical points.  The proof was algebraic in nature and made use of the Euler trace formula.  Our result goes beyond previous work that considered singularities up to codimension five.

\section{Acknowledgments}
\noindent AOP acknowledges the support of NSF Grant DMS-0707003.  Part of this work was conducted at the Petters Research Institute, Belize.

\appendix
\section{Proof of the Main Theorem}
\label{Proof}

\subsection{Overview of the Method and the Euler Trace Formula}
\label{Euler}

\noindent We summarize key elements of our algebraic method; see Paper II for a detailed presentation.  Consider any polynomial $\varphi(x) = a_n x^n + \cdots + a_1 x + a_0 \in \mathbb{C}[x]$ with distinct roots $x_1,\dots,x_n$ and any rational function $h(x)\in R$, where $R \subset \mathbb{C}(x)$ is the subring of rational functions defined at the roots of $\varphi(x)$.  In Paper II we showed algebraically that
for any rational function $h(x) \in R$,  the following holds:
\beqa
\label{euler}
\sum_{i=1}^{n} h(x_i)  = \frac{b_{n-1}}{a_n}\ ,\hspace{.75 in}(\text{Euler Trace Formula})
\eeqa
where 
$b_{n-1}$ is the $(n-1)$st coefficient of the unique polynomial representative $r(x)$ of degree less than $n$ in the coset $\oli{\varphi'(x)\, h(x)} \in \rideal$.  (An alternate proof of the Euler trace formula using residues can be found in Dalal \& Rabin 2001 \cite{Dalal-Rabin}.)  We employ the Euler trace formula as follows.  For any caustic singularity in Table~\ref{table1}, we shall realize the lensed images of its corresponding family of mappings $\bbf_{\bc}$ as solutions of a polynomial in one variable, obtained by eliminating one of the pre-image coordinates, say $y$.  Denote this polynomial by
$$
\varphi(x) = a_n x^n + \cdots + a_1 x + a_0 \in \mathbb{C}[x].
$$
Generically, we can assume that the roots of $\varphi (x)$ are distinct, an assumption made throughout the paper.  We would then be able to express the magnification $\fkM(x,y; \bs)$ at a general pre-image point $(x,y)$ as a function of one variable, in this case $x$, so that
$$
\fkM(x,y(x);\bs) = \frac{1}{\text{det(Jac}\,\bbf_{\bc})(x,y(x))} \equiv \frac{1}{\text{det(Jac}\,\bbf_{\bc})(x)} \equiv \fkM(x)\ ,
$$
where the explicit notational dependence on $\bs$ is dropped for simplicity.  Since we shall consider only non-caustic target points $\bs$ giving rise to lensed images $(x_i, y_i(x_i))$, we have $\text{det(Jac}\,\bbf_{\bc})(x_i) \neq 0$.  We thus know that for the singularities in Table~\ref{table1}, the rational function $\fkM(x)$ is defined at the roots of $\varphi (x)$, i.e., $\fkM(x) \in R$.  Now, denote by $\fkm(x)$ the unique polynomial representative of degree less than $n$ in the coset $\oli{\varphi'(x)\, \fkM(x)} \in \rideal$, and let $b_{n-1}$ be its $(n-1)$st coefficient.  In the notation used above, we have
$h(x) \equiv \fkM(x)$ and 
$r (x) \equiv \fkm(x)$.
Euler's trace formula (Corollary~\ref{euler}) then tells us immediately that the total signed magnification satisfies 
\beq\label{eq:magsum}
\sum_{i} \fkM_i = \frac{b_{n-1}}{a_{n}}\cdot
\eeq
It therefore remains to determine the coefficient $b_{n-1}$ for each caustic singularity in Table~\ref{table1}.

\subsection{Magnification Sum Rule for Type $A_n$}
\label{Amags}

\noindent We begin with type $A_n$, $n \geq 2$.  Since the cases $2 \leq n \leq 6$ were treated in Paper II, we will consider $n \geq 7$ here.  The $(n-1)$-parameter family of general functions $F^{A_n}$ is given in \cite{Arnold73} by
\beq
\label{An1}
F^{A_n}(x,y) = \underbrace{\pm x^{n+1} \pm y^2}_{\text{germ}} + \underbrace{c_{n-1}x^{n-1} + c_{n-2}x^{n-2} + \cdots + c_3x^3 + c_2x^2 + c_1x}_{\text{unfolding terms}}.
\eeq
To convert this into the form shown in Table~\ref{table1}, we use the following coordinate transformation on the domain $\{(x,y)\} = \RR^2$:
\beq
\label{An_coord}
(x,y) \longmapsto \left(x,y+\frac{c_2}{2}\right).
\eeq
This transforms eqn.~\eqref{An1} to
\beq
\label{An}
F_{\bc,\bs}^{A_n}(x,y) = \pm x^{n+1} \pm y^2 + c_{n-1}x^{n-1} + c_{n-2}x^{n-2} + \cdots + c_3x^3 + s_2x^2 - s_1x \pm s_2y\ ,
\eeq
where $c_1 \equiv -s_1$ and $c_2 \equiv s_2$.  The parameters $s_1, s_2$ are to be interpreted in the context of gravitational lensing as the rectangular coordinates on the source plane $S = \RR^2$.  Note that we omitted the constant term from eqn.~\eqref{An} since it will not affect any of our results below.  Note also that
$$
\text{det\big(Hess}\,F^{A_n}\big) = \text{det\big(Hess}\,F_{\bc,\bs}^{A_n}\big)\ ,
$$
so that the magnification (as defined in eqn.~\eqref{Gauss}) is unaltered.  We will work with the form of $F_{\bc,\bs}^{A_n}$ in eqn.~\eqref{An}.  The corresponding $(n-3)$-parameter family of general mappings $\fA\colon\,\mathbb{R}^2 \lra \mathbb{R}^2$ is
\beq
\label{Anmap2}
\fA(x,y) = \left(\pm(n+1)x^{n} + (n-1)c_{n-1}x^{n-2} + (n-2)c_{n-2}x^{n-3} + \cdots + 3c_3x^2 \mp 4yx~,~\mp 2y\right) = (s_1,s_2).\nonumber
\eeq
Here $\bs = (s_1,s_2)$ is a non-caustic target point lying in the region with the maximum number of lensed images.  Since $s_2 = \pm2y$, we can eliminate $y$ to obtain a polynomial in the variable $x$:
\beq
\label{Avar}
\varphi_{A_n}(x) \equiv \pm(n+1)x^{n} + (n-1)c_{n-1}x^{n-2} + (n-2)c_{n-2}x^{n-3} + \cdots + 3c_3x^2 + 2s_2x - s_1\ ,
\eeq
whose $n$ roots are the $x$-coordinates of the lensed images $\bx_i$ of $\bs$.  The Jacobian determinant of $\fA$ expressed in the single variable $x$ is
\beqa
\label{Ajac}
\detfA(x) = \mp2\left[\pm n(n+1)x^{n-1} + (n-2)(n-1)c_{n-1}x^{n-3} + (n-3)(n-2)c_{n-2}x^{n-4} + \cdots + 6c_3x + 2s_2\right].\nonumber\\
\eeqa
A comparison of eqns.~\eqref{Avar} and \eqref{Ajac} then shows that
$$
\pm 2\varphi_{A_n}'(x) = \detfA(x) = \frac{1}{\fkM(x)}\cdot
$$
We thus have
$$
\varphi_{A_n}'(x)\fkM(x) = \pm \frac{1}{2}\cdot
$$
Thus the unique polynomial representative of the coset $\overline{\varphi_{A_n}'(x)\fkM(x)}$ is the polynomial $\fkm(x) \equiv \pm1/2$, whose $(n-1)$st coefficient is $b_{n-1} = 0$ for all $n \geq 7$.  Euler's trace formula in the form of eqn.~\eqref{eq:magsum} then tells us that the total signed magnification is
\beq
\label{Azero}
\sum_{i = 1}^{n} \fkM_i = 0\ ,\hspace{0.2 in} (A_n,\ n\geq 2).\nonumber
\eeq

\subsection{Magnification Sum Rules for Type $D_n$}
\label{Dmags}

\noindent For type $D_n$, $n \geq 4$, the cases $4 \leq n \leq 6$ were treated in Paper II, so we will consider $n \geq 7$ here.  The corresponding $(n-3)$-parameter family of induced general maps $\fD\colon\,\mathbb{R}^2 \lra \mathbb{R}^2$ is shown in Table~\ref{table1}:
\beq
\label{Dnmap2}
\fD(x,y) = \left(2xy~,~x^2\pm(n-1)y^{n-2} + (n-2)c_{n-2}y^{n-3} + \cdots + (n-i)c_{n-i}y^{n-(i+1)} + \cdots + 2c_2y\right) = (s_1,s_2).
\eeq
Once again the point $\bs = (s_1,s_2)$ is a non-caustic target point lying in the region with the maximum number of lensed images.  This time, however, we eliminate $x$ to obtain a polynomial in the variable $y$:
$$
\varphi_{D_n^{\pm}}(y) \equiv \pm 4(n-1)y^{n} + 4(n-2)c_{n-2}y^{n-1} + \cdots + 4(n-i)c_{n-i}y^{n-(i-1)} + \cdots + 8c_2y^{3} - 4s_2y^2 + s_1^2\ ,
$$  
whose $n$ roots are the $y$-coordinates of the $n$ lensed images $\bx_i$ of $\bs$.  The derivative of $\varphi_{D_n^{\pm}}(y)$ is 
\beq
\label{Dder}
\varphi_{D_n^{\pm}}'(y) = \pm 4n(n-1)y^{n-1} + 4(n-1)(n-2)c_{n-2}y^{n-2} + \cdots + 4(n-(i-1))(n-i)c_{n-i}y^{n-i} + \cdots + 24c_2y^{2} - 8s_2y\ ,
\eeq
while the Jacobian determinant of $\fD$ is
\beqa
\label{Dvar}
\detfD(x,y) &=& \text{det}\!\left[\begin{array}{cc} 2y & 2x \\ 2x & \pm(n-2)(n-1)y^{n-3} + (n-3)(n-2)c_{n-2}y^{n-4} + \cdots + 2c_2 \\ \end{array}\right]\nonumber \\
&=& \pm2(n-2)(n-1)y^{n-2} + 2(n-3)(n-2)c_{n-2}y^{n-3} + \cdots \nonumber\\
&&\cdots + 2(n-(i+1))(n-i)c_{n-i}y^{n-(i+1)} + \cdots + 4c_2y - 4x^2.\nonumber
\eeqa
We can use eqn.~\eqref{Dnmap2} to eliminate $x$ as follows:
\beqa
&=& \pm 2(n-2)(n-1)y^{n-2} + 2(n-3)(n-2)c_{n-2}y^{n-3} + \cdots + 2(n-(i+1))(n-i)c_{n-i}y^{n-(i+1)} + \cdots + 4c_2y\nonumber \\
& &+4\underbrace{\left(\pm(n-1)y^{n-2} + (n-2)c_{n-2}y^{n-3} + \cdots + (n-i)c_{n-i}y^{n-(i+1)} + \cdots + 2c_2y - s_2\right)}_{=\,-x^2~~(\text{by eqn.}~\eqref{Dnmap2})}\nonumber \\
&=& \pm2n(n-1)y^{n-2} + 2(n-1)(n-2)c_{n-2}y^{n-3} + \cdots + 2(n-(i-1))(n-i)c_{n-i}y^{n-(i+1)} + \cdots + 12c_2y - 4s_2 \nonumber \\
&=& \detfD(y) = \fkM(y)^{-1}.\nonumber
\eeqa
A comparison with eqn.~\eqref{Dder} then shows that
$$
\varphi_{D_n^{\pm}}'(y)\fkM(y) = 2y.
$$
The unique polynomial representative of the coset $\overline{\varphi_{D_n^{\pm}}'(y)\fkM(y)}$ is therefore the polynomial $\fkm(y) \equiv 2y$, whose $(n-1)$st coefficient is $b_{n-1} = 0$ for all $n \geq 7$.  Eqn.~\eqref{eq:magsum} then tells us that the total signed magnification is
\beq
\label{Dzero}
\sum_{i = 1}^{n} \fkM_i = 0\ ,\hspace{0.2 in} (D_n,\ n\geq 4).\nonumber
\eeq

\subsection{Magnification Sum Rules for Types $E_n$}
\label{Emags}

\noindent The case $E_6$ corresponds to the symbolic umbilic, whose magnification sum rule was proved in Paper II.  For type $E_7$, Table~\ref{table1} gives the corresponding 4-parameter family of induced general maps $\fEseven\colon\,\mathbb{R}^2 \lra \mathbb{R}^2$:
\beq
\label{E7map2}
\fEseven(x,y) = \left(3x^2 + y^3 + c_1y~,~3xy^2 + 4c_4y^3 + 3c_3y^2 + 2c_2y + c_1x\right) = (s_1,s_2).
\eeq
Once again, the point $\bs = (s_1,s_2)$ is a non-caustic point lying in the region with the maximum number of lensed images.  We eliminate $x$ to obtain a polynomial in the variable $y$:
\beqa
\varphi_{E_7}(y) &\equiv& 9y^7 + 48c_4^2 y^6 + \left(15c_1 + 72c_3 c_4\right)y^5 + \left(27 c_3^2 + 48 c_2 c_4 - 9 s_1\right)y^4 + \left(7 c_1^2 + 36 c_2 c_3 - 24 c_4 s_2\right) y^3\nonumber \\
&& +\ \left(12 c_2^2 - 6 c_1 s_1 - 18 c_3 s_2\right) y^2 + \left(c_1^3 - 12 c_2 s_2\right)y -c_1^2 s_1 + 3 s_2^2\ ,\nonumber
\eeqa
whose 7 roots are the $y$-coordinates of the 7 lensed images $\bx_i$ of $\bs$.  The derivative of $\varphi_{E_7}(y)$ is 
\beqa
\label{derseven}
\varphi_{E_7}'(y) &=& 63y^6 + 288c_4^2 y^5 + \left(75c_1 + 360c_3c_4\right)y^4 + \left(108c_3^2 + 192c_2 c_4 - 36s_1\right)y^3\nonumber \\
&& +\ \left(21c_1^2 + 108c_2 c_3 - 72c_4 s_2\right) y^2 + \left(24c_2^2 - 12c_1 s_1 - 36c_3 s_2\right)y + c_1^3 - 12 c_2 s_2\ ,
\eeqa
while the Jacobian determinant of $\fEseven$ is 
\beqa
\detfEseven(x,y) &=& \text{det}\!\left[\begin{array}{cc} 6x & 3y^2 + c_1 \\ 3y^2 + c_1 & 6xy + 12c_4y^2 + 6c_3y + 2c_2 \\ \end{array}\right]\nonumber \\ 
&=& 36x^2y + 72c_4xy^2 + 36c_3xy + 12c_2x -9y^4 - 6c_1y^2 - c_1^2.\nonumber
\eeqa
To convert this into a function in the single variable $y$, we use eqn.~\eqref{E7map2} twice, as follows:
\beqa
\label{E7mag2}
&=& 36x^2y + 72c_4xy^2 + 36c_3xy + 12c_2x -9y^4 - 6_1y^2 - c_1^2\nonumber \\
&=& 36y\underbrace{\left(\frac{s_1 -c_1y - y^3}{3}\right)}_{=\,x^2~~(\text{by eqn.}~\eqref{E7map2})} + \left(72c_4y^2 + 36c_3y + 12c_2\right)\underbrace{\left(\frac{s_2 - 4c_4y^3 - 3c_3y^2 - 2c_2y}{3y^2 + c_1}\right)}_{=\,x~~(\text{by eqn.}~\eqref{E7map2})} -9y^4 - 6c_1y^2 - c_1^2\nonumber \\
&\vdots&\nonumber\\
&=&  \left(c_1 + 3 y^2\right)^{-1} \big[-c_1^3 + 12 c_2 s_2 + \left(-24 c_2^2 + 12 c_1 s_1 + 36 c_3 s_2\right)y + \left(-21 c_1^2 - 108 c_2 c_3 + 72 c_4 s_2\right)y^2\big.\nonumber \\
&&\big. + \left(-108 c_3^2 - 192 c_2 c_4 + 36 s_1\right)y^3 - \left(75 c_1 + 360 c_3 c_4\right)y^4 - 288 c_4^2 y^5 - 63 y^6\big]\nonumber \\
&=& \detfEseven(y) = \fkM(y)^{-1}.
\eeqa
A comparison of eqns.~\eqref{derseven} and \eqref{E7mag2} then shows that
$$
\varphi_{E_7}'(y)\fkM(y) = -(c_1 + 3y^2)\ ,
$$
and so the unique polynomial representative of the coset $\overline{\varphi_{E_7}'(y)\fkM(y)}$ is the polynomial $\fkm(y) \equiv -(c_1 + 3y^2)$, whose 6th coefficient is $b_{6} = 0$.  We conclude via eqn.~\eqref{eq:magsum} that the total signed magnification is
$$
\sum_{i = 1}^{7} \fkM_i = 0\ ,\hspace{0.2 in} (E_7).
$$

\noindent For type $E_8$, Table~\ref{table1} gives the corresponding 5-parameter family of induced general maps $\fEeight\colon\,\mathbb{R}^2 \lra \mathbb{R}^2$:
\beq
\label{E8map2}
\fEeight(x,y) =  \left(3x^2 + c_5y^3 + c_4y^2 + c_1y~,~5y^4 + 3c_5xy^2 + 2c_4xy + 3c_3y^2 + 2c_2y + c_1x\right) = (s_1,s_2).
\eeq
As usual, the point $\bs = (s_1,s_2)$ is a non-caustic target point lying in the region with the maximum number of lensed images.  We eliminate $x$ to obtain a polynomial in the variable $y$,
\beqa
\varphi_{E_8}(y) &\equiv& 75y^8 + 9c_5 y^7 + \left(90 c_3 + 21 c_4 c_5^2\right)y^6 + \left(60 c_2 + 16 c_4^2 c_5 + 15 c_1 c_5^2\right)y^5 + \left(27 c_3^2 + 4 c_4^3 + 22 c_1 c_4 c_5 - 9 c_5^2 s1 - 30 s_2\right)y^4\nonumber \\
&&+ \left(36 c_2 c_3 + 8 c_1 c_4^2 + 7 c_1^2 c_5 - 12 c_4 c_5 s_1\right) y^3 +\ \left(12 c_2^2 + 5 c_1^2 c_4 - 4 c_4^2 s_1 - 6 c_1 c_5 s_1 - 18 c_3 s_2\right) y^2\nonumber \\
&& + \left(c_1^3 - 4 c_1 c_4 s_1 - 12 c_2 s_2\right)y -c_1^2 s_1 + 3 s_2^2\ ,\nonumber
\eeqa
whose 8 roots are the $y$-coordinates of the 8 lensed images $\bx_i$ of $\bs$.  The derivative of $\varphi_{E_8}(y)$ is 
\beqa
\label{dereight}
\varphi_{E_8}'(y) &=& 600y^7 + 63c_5 y^6 + \left(540c_3 + 126c_4 c_5^2\right)y^5 + \left(300 c_2 + 80 c_4^2 c_5 + 75 c_1 c_5^2\right)y^4\nonumber \\
&&+ \left(108 c_3^2 + 16 c_4^3 + 88 c_1 c_4 c_5 - 36 c_5^2 s_1 - 120 s_2\right)y^3 +\ \left(108 c_2 c_3 + 24 c_1 c_4^2 + 21 c_1^2 c_5 - 36 c_4 c_5 s_1\right) y^2\nonumber \\
&&+ \left(24 c_2^2 + 10 c_1^2 c_4 - 8 c_4^2 s_1 - 12 c_1 c_5 s_1 - 36 c_3 s_2\right)y + c_1^3 - 4 c_1 c_4 s_1 - 12 c_2 s_2\ ,
\eeqa
and the Jacobian determinant of $\fEeight$ is 
\beqa
\label{E8mag}
\detfEeight(x,y) &=& \text{det}\!\left[\begin{array}{cc} 6x & 3c_5y^2 + 2c_4y + c_1 \\ 3c_5y^2 + 2c_4y +c_1 & 20y^3 + 6c_5xy + 2c_4x + 6c_3y +2c_2 \\ \end{array}\right]\nonumber \\ 
&=& -c_1^2 + 12 c_2 x + 12 c_4 x^2 - 4 c_1 c_4 y + 36 c_3 x y + 36 c_5 x^2 y - 4 c_4^2 y^2\nonumber \\
&& -6 c_1 c_5 y^2 - 12 c_4 c_5 y^3 + 120 x y^3 - 9 c_5^2 y^4.\nonumber
\eeqa
Similar to the case $E_7$ above, we convert this to a function in the single variable $y$ with the aid of eqn.~\eqref{E8map2}:
\beqa
\label{M8}
&=& -c_1^2 + 12 c_2 x + 12 c_4 x^2 - 4 c_1 c_4 y + 36 c_3 x y + 36 c_5 x^2 y - 4 c_4^2 y^2\nonumber \\
&& -6 c_1 c_5 y^2 - 12 c_4 c_5 y^3 + 120 x y^3 - 9 c_5^2 y^4\nonumber\\
&=& \left(12c_2 + 36c_3y + 120y^3\right)\underbrace{\left(\frac{s_2 - 5y^4 - 3c_3y^2 - 2c_2y}{3c_5y^2 + 2c_4y +c_1}\right)}_{=\,x~~(\text{by eqn.}~\eqref{E8map2})} + \left(12c_4 + 36c_5y\right)\underbrace{\left(\frac{s_1 - c_1y - c_4y^2 - c_5y^3}{3}\right)}_{=\,x^2~~(\text{by eqn.}~\eqref{E8map2})}\nonumber\\
&& - c_1^2 - 4c_1c_4y - 4c_4^2y^2 - 6c_1c_5y^2 - 12c_4c_5y^3 - 9c_5^2y^4\nonumber\\
&\vdots&\nonumber\\
&=&  \left(c_1 + 2c_4 y + 3c_5 y^2\right)^{-1} \big[-c_1^3 + 4 c_1 c_4 s_1 + 12 c_2 s_2 +\left(- 24 c_2^2 - 10 c_1^2 c_4 + 8 c_4^2 s_1 + 12 c_1 c_5 s_1 + 36 c_3 s_2\right)y\nonumber \\
&& +\left(- 108 c_2 c_3 - 24 c_1 c_4^2 - 21 c_1^2 c_5 + 36 c_4 c_5 s_1\right) y^2 +\left(- 108 c_3^2 - 16 c_4^3 - 88 c_1 c_4 c_5 + 36 c_5^2 s_1 + 120 s_2\right)y^3\nonumber \\
&& +\left(- 300 c_2 - 80 c_4^2 c_5 - 75 c_1 c_5^2\right)y^4 -\left(540 c_3 + 126 c_4 c_5^2\right)y^5 - 63 c_5^3 y^6 - 600 y^7\big]\nonumber\\
&=& \detfEeight(y)= \fkM(y)^{-1}.
\eeqa
A comparison of eqns.~\eqref{dereight} and \eqref{M8} thus shows that
$$
\varphi_{E_8}'(y)\fkM(y) = -(c_1 + 2c_4 y + 3c_5 y^2)\ ,
$$
and so the unique polynomial representative of the coset $\overline{\varphi_{E_8}'(y)\fkM(y)}$ is the polynomial $\fkm(y) \equiv -(c_1 + 2c_4 y + 3c_5 y^2)$, whose 7th coefficient is $b_{7} = 0$.  Euler's trace formula tells us once again that
$$
\sum_{i = 1}^{8} \fkM_i = 0\ ,\hspace{0.2 in} (E_8).
$$
This completes the proof of the magnification relations for {\it all} higher-order caustic singularities appearing in the infinite $A_n~(n \geq 2), D_n~(n \geq 4), E_6, E_7, E_8$ family. $\qed$

\newpage

\end{document}